\documentclass[twocolumn]{aa}

\usepackage{longtable}
\usepackage{graphicx,caption}

\usepackage{txfonts}
\usepackage{longtable,lscape}

\def\R23{\mbox{$\rm R_{23}$}}

\def\kmsmpc{km s$^{-1}$ Mpc$^{-1}$}

\def\msun{M$_{\odot}$}

\def\Hb{\mbox{${\rm H}{\beta}$}}
\def\Ha{\mbox{${\rm H}{\alpha}$}}

\def\OIIIa{\mbox{${\rm [O\,III]\,}{\lambda\,5007}$}}

\def\OIIa{\mbox{${\rm [O\,II]\,}{\lambda\,3726}$}}
\def\OIIb{\mbox{${\rm [O\,II]\,}{\lambda\,3729}$}}
\def\OIId{\mbox{${\rm [O\,II]\,}{\lambda\lambda\,3726,3729}$}}

\def\NII{\mbox{${\rm [N\,II]\,}{\lambda\,6584}$}}

\begin{document}

\title{The BLOBs: Enigmatic diffuse ionized gas structures  in a cluster of galaxies near cosmic noon}

\author{C.~Maier\inst{1}
\and B.~L.~Ziegler\inst{1}
\and T.~Kodama\inst{2}
}

\institute{University of Vienna, Department of Astrophysics, Tuerkenschanzstrasse 17, 1180 Vienna, Austria\\
\email{christian.maier@univie.ac.at}
\and Astronomical Institute, Tohoku University, Aramaki, Aoba-ku, Sendai 980-8578, Japan
}

\titlerunning{Oxygen blobs and merging in a cluster at $z \sim 1.5$}
\authorrunning{C. Maier et al.}

\date{Received ; accepted}

\abstract
{
We explore the massive cluster XMMXCS\,J2215.9-1738 at $z \sim 1.46$ with MUSE and KMOS integral field spectroscopy.
Using MUSE spectroscopy, we traced the kinematics of the ionized gas using \OIId\, in the central 500$\times$500\,kpc$^{2}$  area of the cluster, which contains 28 spectroscopically identified cluster galaxies. We detected \OIId\, emission lines in the integrated spectra of 21 galaxies. The remaining seven are passive galaxies. 
Six of these passive galaxies lie in the central part of the cluster, which has a diameter of 200\,kpc and contains no star-forming objects. In this place, star formation in galaxies is quenched. 
An interesting discovery in this central area of the cluster are three diffuse ionized \OIId\, gas structures, which we refer to as [OII] blobs. They extend over areas of some hundred kpc$^{2}$.
The ionization source of one of the gaseous structures that displays two prominent filamentary patterns indicating outflow of gas is an active galactic nucleus (AGN).
The KMOS data enabled us to
use the BPT diagram to identify this object as a type 2 AGN.
The other two diffuse ionized oxygen gaseous structures
are more enigmatic. They are located between the stellar components of passive cluster galaxies.
One of these blobs lacks a stellar counterpart in the HST optical and near-infrared data, and the other blob has only a very faint counterpart.
Ram-pressure stripping of photoionized gas or AGN feedback
might be an explanation. Additionally, the galaxy velocity distribution in this high-redshift cluster is bimodal, which indicates that the cluster is probably not fully virialized and that recent and ongoing merging events that produced shocks might provide photoionization sources for the two enigmatic [OII] blobs.
}

\keywords{
  Galaxies: evolution -- Galaxies: clusters: general-- Galaxies: star formation--  Galaxies: clusters: XMMXCS\,J2215.9-1738
}

\maketitle

\setcounter{section}{0}

\section{Introduction}
\label{sec:intro}
 
~~~The environment plays a major role in shaping galaxy evolution. Rich clusters of galaxies in the local and intermediate-redshift Universe are dominated by a quiescent galaxy population composed of ellipticals and lenticulars \citep[e.g.,][]{dressler80,dressler97}. The identification of the dominant physical mechanism that removes the gas and subsequently quenches the star formation activity in cluster galaxies is one of the major questions in galaxy evolution. Different physical processes have been proposed in the literature to explain this (see the review by \citet{bosgav14}).  These processes can be broadly divided into two main classes: Processes related to the gravitational perturbation of a galaxy in a dense environment (galaxy–galaxy interactions, galaxy-cluster interactions, or galaxy harassment), and processes due to the hydrodynamical interaction of the interstellar medium with the hot and dense intracluster medium (ram pressure, thermal evaporation, viscous stripping, and starvation). Purely hydrodynamical interactions such as ram pressure leave the stellar body of a galaxy almost undisturbed and only affect its gaseous components, while gravitational interactions act on both.

~~~While local and low-redshift clusters host large fractions of passive galaxies, the star-forming (SF) population in clusters becomes larger with increasing redshift. This enables a study of the SF cluster galaxies. Since galaxy clusters grow by accreting mass from their surroundings, the SF galaxies observed in high-z clusters must be the progenitors of the local passive galaxies. The $1 < z < 2$ redshift range is a transition epoch for clusters and hosts the emergence of the Hubble
sequence of disks and elliptical galaxies  and the buildup of a
significant fraction of the stellar mass in the Universe \citep[e.g.,][]{dickins03,drory05}.
For our studies of environmental effects  at $1 < z < 2$, we selected the $z \sim 1.46$ cluster XMMXCS\,J2215.9-1738 (hereafter XMM2215). One advantage of observational environmental studies in this cluster compared to other clusters at $z>1$ is its wealth of ancillary data \citep[see][M19 in the following]{maier19}.

~~~ Diffuse \OIId\, ionized gas traces warm 10\,000\,K gas. \citet{yuma13} introduced a first systematic search
for galaxies with spatially extended \OIId\, emission (called [OII] blobs). They found that these blobs are large up to 75\,kpc and are mainly associated with AGN outflows. In a later publication,  \citet{yuma17} exploited the number density of [OII] blobs further and reported that it decreases drastically at $z<1.5$ with redshift at a rate that is higher than that of the decrease in the cosmic star formation density.
At lower redshifts, a serendipitous discovery of a large ionized [OII] gas structure observed in an overdense region of a galaxy group at redshift $z\sim 0.7$  was reported by \citet{epinat18}. In the local Universe, the compact group called Stephan's Quintet was found to contain a 35 kpc intergalactic filamentary structure that was observed in radio continuum as well as in optical emission lines (ELs) and X-rays \citep{baras14}.

\begin{figure}[!t]
  \centering
  \captionsetup{width=0.45\textwidth}
    \includegraphics[width=0.45\textwidth,angle=0,clip=true]{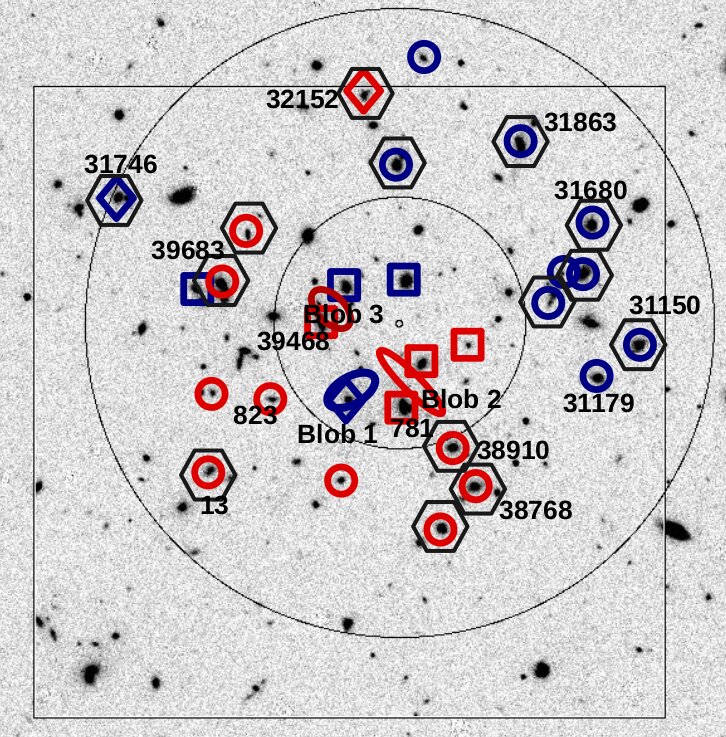}
   \centering   
   \caption{\footnotesize HST F160W image with the distribution of the 28 cluster members with spectroscopic redshifts in the center of the XMM2215 cluster. The 500$\times$500\,kpc$^{2}$  area (at $z\sim 1.46$) covered by MUSE is shown by the large black square. Only the 28 galaxies are shown that lie inside a circular area (indicated by the large black circle) with a radius of 250\,kpc around the cluster X-ray center (tiny open black circle).  Fourteen galaxies with $z<1.457$ (and 14 at $z>1.457$) are highlighted in blue (and red); SF galaxies by open circles, three AGN  by  blue and red rhombuses, and passive galaxies by blue and red squares. The IDs follow the notation that was used by M19. The 14 galaxies with ALMA molecular gas measurements are in addition denoted by black hexagons. The rough positions and form of the three blobs of diffuse warm ionized [OII] gas are shown by blue and red ellipses (the color depends on the redshift). The smaller black circle with a diameter of $\sim$200\,kpc around the cluster center is populated only by passive galaxies, an AGN, and the ionized gas blobs. The galaxies with ALMA molecular gas detections (hexagons) all lie outside this area.    
   }  
\label{fig:clustcen}
\end{figure}

~~~We present our investigation of the warm ionized gas in cluster XMM2215 using Multi Unit Spectroscopic Explorer (MUSE) and K-band Multi Object Spectrograph (KMOS) integral field spectroscopic observations at the Very Large Telescope (VLT). The measurement of \OIId\, ELs allowed us to study the kinematics and to identify  [OII] blobs at $z \sim 1.46$.

~~~The paper is structured as follows: In Sect.\,\ref{sec:obs}  we
describe
the data reduction of the MUSE and KMOS observations and the measurement of the EL fluxes.
In Sect.\,3 we present our results, including the bimodal galaxy velocity distribution in the cluster center, the detection of three [OII] blobs, and the properties of the ionized gas of other SF galaxies in the central area of the cluster.
In Sect.\,4 
we discuss possible ionization sources for the three [OII] blobs, possible reasons for the
quenching of galaxies in the cluster center, and a particular merger event traced by the warm ionized \OIId\, gas. We also address caveats and next steps.
Finally, we summarize our conclusions.
A concordance cosmology with $\rm{H}_{0}=70$ \kmsmpc,
$\Omega_{0}=0.25$, $\Omega_{\Lambda}=0.75$ is used throughout this
paper. 

\section{Observations, data reduction, and flux measurements}
\label{sec:obs}

~~~The cluster XMM2215 is a massive $z \sim 1.46$ cluster that was discovered in the XMM Cluster Survey \citep{stanford06}, with extended X-ray emission from the hot gas that suggests that the cluster is in a relatively advanced evolutionary stage. Estimates of its mass and $R_{200}$ vary from $M_{200} = 2.1^{-0.8}_{+1.9} \cdot 10^{14}M_{\odot}$ and $R_{200} = (0.8 \pm 0.1)$Mpc \citep{hilton10,stott10} to  $M_{200} = (6.3 \pm 1.2) \cdot 10^{14}M_{\odot}$ and $R_{200} = (1.23 \pm 0.18)$Mpc (M19). However, the cluster is probably not fully virialized because the galaxy velocity distribution is bimodal (see Sect.\,\ref{sec:bimodalvel}), and there is no clear brightest cluster galaxy \citep{hilton09,stott10}.  The whole wealth of archival ancillary data for this cluster has recently been summarized in detail by \citet{klutse24} (we refer to this for details). We focus  on a description of the XMM2215 data that are most relevant for our study, which we reduced and analyzed ourselves, namely the MUSE and KMOS data.

\begin{figure*}[!t]
  \centering
  \captionsetup{width=0.95\textwidth}
    \includegraphics[width=0.85\textwidth,angle=0,clip=true]{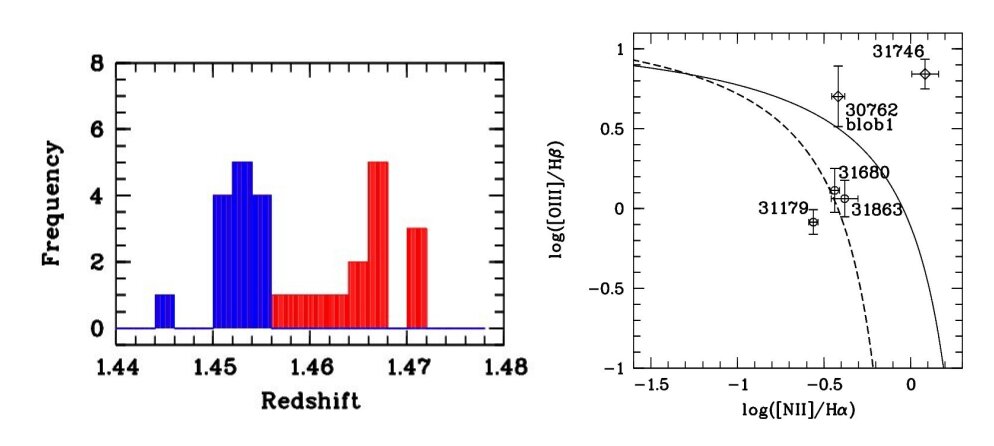}
   \centering   
   \caption{\footnotesize Left panel: Bimodal distribution for the redshifts of 28 cluster galaxies 
     inside a circular area,  shown by the large black circle in Fig.\,\ref{fig:clustcen}, with a radius of 250\,kpc around the cluster X-ray center.
     Right panel: Diagnostic diagram BPT \citep{bald81} to distinguish star formation dominated galaxies from AGN for objects with all four observed ELs. The SF region of the BPT diagram lies below and to  the left of the empirical (dashed) curve of \citet{kaufm03}, while the composite region lies between the theoretical (solid) curve of \citet{kew01} and the curve of \citet{kaufm03}.  Two XMM2215 cluster galaxies are classified as type 2 AGN according to this diagram. Two galaxies lie in the composite region, and one galaxy lies in the region dominated by star formation. Because KMOS IFUs were only placed on galaxies with stellar counterparts detected on images, which was not the case for blob\,2 and blob\,3, only blob\,1 (object 30762) is depicted in the BPT diagram. Additionally, as further discussed in Sect.\,\ref{sec:caveats}, during the VLT-KMOS observations the night skylines would have affected some of the ELs needed for the BPT diagram for blob\,2 and blob\,3 strongly, given their redshifts. 
   }  
\label{fig:zHist}
\end{figure*}

\begin{figure*}[!t]
  \centering
  \captionsetup{width=0.95\textwidth}
    \includegraphics[width=0.95\textwidth,angle=0,clip=true]{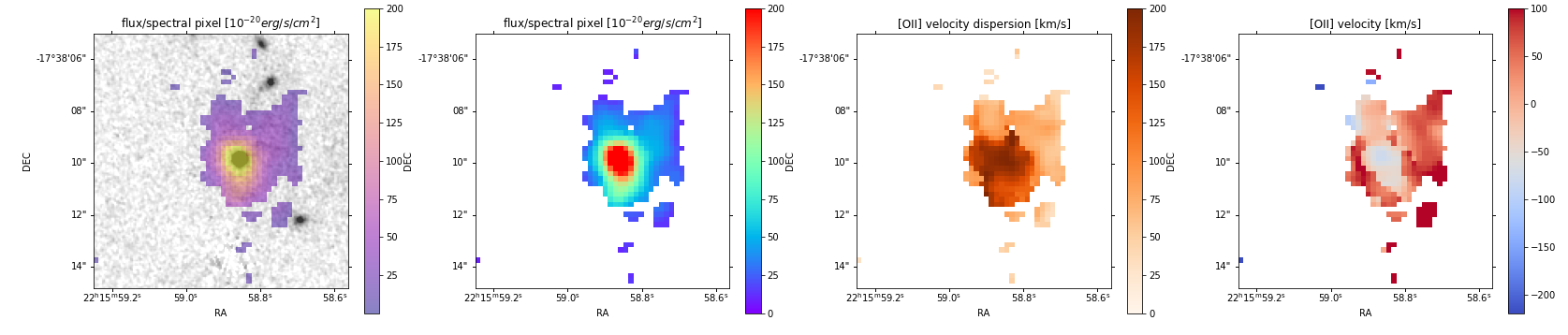}
   \centering   
   \caption{\footnotesize From left to right panel:  HST F160W image around galaxy 30762 (blob\,1) with the [OII] EL map overlaid, the spatially resolved [OII] EL map, the spatially resolved [OII] velocity dispersion map, and the  spatially resolved [OII] velocity map. The white background corresponds to spaxels with $\rm{S/N_{[OII]}<5}$.
   }  
\label{fig:AGN30762}
\end{figure*}

~~~Observations of the central part of XMM2215 were taken with the MUSE integral field unit using the wide-ﬁeld adaptive optics (WFM-AO) mode as part of ESO programme 60.A-9180. In total, the cluster was observed for about 7ks, split into eight 864 s exposures, using the standard spectral range covering $4770-9300$\AA\, and a spectral resolution of $R=\lambda/\Delta \lambda =4000$ at $\lambda=9200$\AA. We used the raw data from the  ESO Science Archive Facility and reduced them using the standard calibrations provided by the ESO-MUSE pipeline, version 1.2.1 \citep{pipeline}.  To reduce the sky residuals, we used an additional tool,  the \texttt{Zurich Atmosphere Purge version 1.0} (\texttt{ZAP}, \citealt{soto}), on the calibrated cube.

~~~To derive [OII] EL flux and kinematics maps, we first extracted for every cluster galaxy and blob a subcube of 50$\times$50 spatial spaxels, corresponding to 10$\times$10 arcsec$^{2}$.
We used the Python code \emph{CAMEL} described by \citet{epinat12} to extract the ionized gas kinematics of galaxies and blobs by fitting ELs in subcubes with a $10 \times 10$ arcsec$^2$ field of view (FoV) around the galaxies or blobs. Before we extracted these maps, a spatial smoothing using a 2D Gaussian with a FWHM of 3 pixels was applied to each data subcube in order to increase the signal-to-noise ratio (S/N). For each spaxel, the [OII] doublet was modeled with two Gaussian profiles that shared the same kinematics (same velocity and same velocity dispersion), but had distinct rest-frame wavelengths (3726.04 \AA\ and 3728.80 \AA) plus a constant continuum. The variance data cube was used to weight each spectral element during the line fitting in order to minimize the effect of noise, which is mainly due to the sky lines.
For the three blobs, we considered [OII] fluxes and radial velocities for spaxels with [OII] EL fluxes with an $S/N>5$.

~~~We concentrated on the inner part of the cluster that is almost completely covered by the MUSE data (see Fig.\,\ref{fig:clustcen}).  With the assumed cosmology, 1 arcsec corresponds to 8.6\,kpc at the mean redshift of the cluster. Thus, the extent of the MUSE FoV
is equivalent to 500$\times$500\,kpc$^{2}$ at $z\sim 1.46$ in the cluster center.

~~~The KMOS H-band observations used to measure the \Ha\, and \NII\, ELs of $z \sim 1.46$ cluster galaxies were described in M19. The VLT/KMOS YJ-band observations aimed to measure the \OIIIa\, and \Hb\, ELs were carried out in September 2022 (ESO program ID 109.22VG) with a nod-to sky strategy and each observing block (OB) consisting of  several ABA ABA sequences. The integration time of each exposure was 420\,s, yielding a total exposure time of 16800\,s. The KMOS data for the YJ-data were reduced with the official ESO-KMOS pipeline with the same steps as described in M19 for the KMOS H-band data.
To measure the EL flux ratios \NII/\Ha\, and \OIIIa/\Hb\, needed for the BPT \citep{bald81} diagram,  we extracted 1D spectra summed over 25 KMOS spaxels centered on the  \Ha\, (\OIIIa\,) EL of the respective H-band or YJ-band KMOS data cube (see more details in M19). This corresponds to an aperture of 1 square arcsecond or 8.6$\times$8.6\,kpc$^{2}$ at these redshifts.

\section{Results}
\label{sec:results}

\begin{figure*}[!h]
  \centering
  \captionsetup{width=0.9\textwidth}
    \includegraphics[width=0.9\textwidth,angle=0,clip=true]{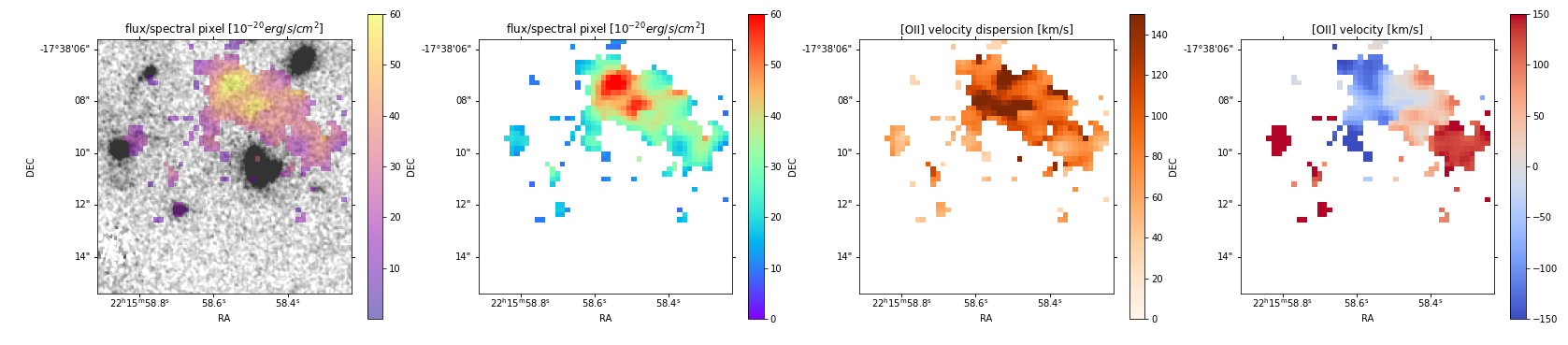}
   \centering   
   \caption{\footnotesize From left to right panel:  HST F160W image including blob\,2 with the [OII] EL map overlaid, the spatially resolved [OII] EL map, the spatially resolved [OII] velocity dispersion map, and the  spatially resolved [OII] velocity map. The white background corresponds to spaxels with $\rm{S/N_{[OII]}<5}$.
   }  
\label{fig:Blob2}
\end{figure*}

\begin{figure*}[!t]
  \centering
  \captionsetup{width=0.95\textwidth}
    \includegraphics[width=0.95\textwidth,angle=0,clip=true]{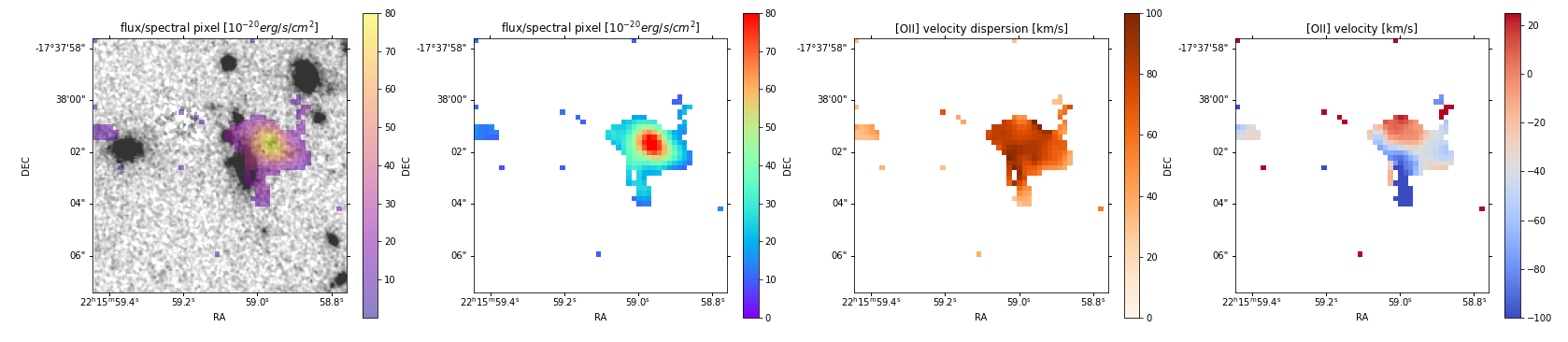}
   \centering   
   \caption{\footnotesize From left to right panel:  HST F160W image including blob\,3 with the [OII] EL map overlaid, the spatially resolved [OII] EL map, the spatially resolved [OII] velocity dispersion map, and the  spatially resolved [OII] velocity map. The white background corresponds to spaxels with $\rm{S/N_{[OII]}<5}$.
   }  
\label{fig:Blob3}
\end{figure*}

\begin{figure*}[!t]
  \centering
  \captionsetup{width=0.95\textwidth}
    \includegraphics[width=0.95\textwidth,angle=0,clip=true]{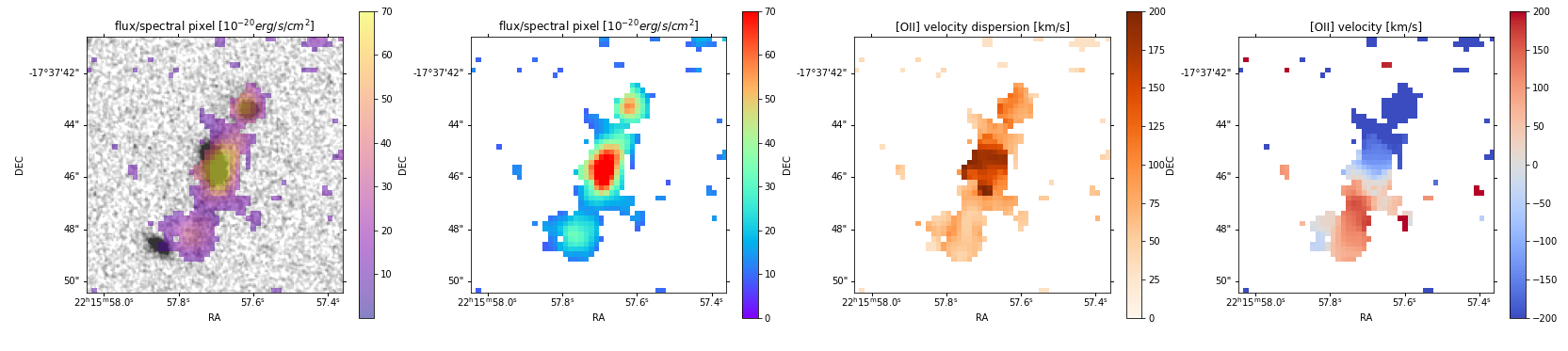}
   \centering   
   \caption{\footnotesize From left to right panel:  HST F160W image around the galaxy 31863 with the [OII] EL map overlaid, the spatially resolved [OII] EL map, the spatially resolved [OII] velocity dispersion map, and the  spatially resolved [OII] velocity map. The white background corresponds to spaxels with $\rm{S/N_{[OII]}<4}$.
   }  
\label{fig:31863}
\end{figure*}

\subsection{General properties of the XMM2215 cluster galaxies}
\label{sec:bimodalvel}

~~~Fig.\,\ref{fig:clustcen} depicts the HST F160W image with the distribution of the 28 cluster members with spectroscopic redshifts in the center of the XMM2215 cluster.  Fourteen galaxies with spectroscopic $z<1.457$ and 14 at $z>1.457$ are highlighted in blue and red. The inner black circle in Fig.\,\ref{fig:clustcen} comprises the central part of the cluster and has a diameter of 200\,kpc. It contains six passive galaxies with known spectroscopic redshifts \citep{hilton10,beifiori17,hayashi18} and one AGN, galaxy 30762, and the corresponding blob\,1 (blue rhombus and ellipse).
ALMA band 3, 6 and 7  observations \citep{hayashi17,stach17,hayashi18} discovered molecular gas for the 14 SF cluster galaxies that are shown as black hexagons in  Fig.\,\ref{fig:clustcen}, and these galaxies with molecular gas detections all lie outside the inner area of the cluster
with a diameter of 200\,kpc.

~~~The histogram of spectroscopic redshifts shown in the left panel of Fig.\,\ref{fig:zHist} indicates a bimodal velocity distribution inside the inner cluster area with a diameter of 500\,kpc that corresponds to a circular area with a radius of about 0.25$R_{200}$ (large black circle in Fig.\,\ref{fig:clustcen}).
Additionally, the 2D distribution of objects in Fig.\,\ref{fig:clustcen} shows a clear separation in the plane of the sky that corresponds to two kinematically distinct structures of galaxies with $z>1.457$ (red symbols) in the southeast region of the HST image and $z<1.457$ (blue symbols) in the northwest region.
With the additional cluster spectroscopic data  obtained  since 2010, we thus find stronger evidence for this effect. This places the mild evidence for a bimodal velocity distribution found by \citet{hilton10} on a stronger footing.

\subsection{Blob\,1: Gas in and around galaxy 30762, which hosts an AGN}

~~~Blob\,1 at  $z \sim 1.453$ at a projected distance of 74\,kpc from the cluster X-ray center (see the table in M19) has a surface of 1145\,kpc$^{2}$, is located at and around the position of the cluster galaxy 30762, and has EL flux measurements from MUSE and KMOS.
The surface area was determined by taking only spaxels with an [OII] flux with an S/N ratio higher than five into account.

~~~The left panel of Fig.\,\ref{fig:AGN30762} shows the position of the stellar component of galaxy 30762 from HST F160W data compared to the ionized oxygen gas.
The MUSE and HST datasets are generally well aligned in terms of the centroid, as seen not only for galaxy 30762, but also for other galaxies with more regular [OII] rotation pattern (see the figures for objects 31179, 31680, 31150, and 32152 in the appendix\,\ref{a1}). We used the image registration Python package, and we measured an astrometric offset between the MUSE data and HST of less than 0.2" (the size of a MUSE spaxel).
The ionized gas of this blob extends well beyond the visible stellar component of the galaxy and exhibits two filamentary patterns, one in the north,  and the other in the northwest direction. The north filament has a projected receding velocity  of $\sim 100$\,km/s according to the [OII] velocity map, and the  northwest filament has  a receding velocity of $\sim 175$\,km/s compared to the blueish area in the center of the blob.

~~~This galaxy was not detected by the ALMA search for molecular gas \citep{hayashi17,stach17,hayashi18}, and not by MeerKAT L-band (1.3 GHz) observations with an integration time of about 12\,hours either \citep{klutse24}.
Using Chandra observations of the cluster, \citet{hilton10} identified object 30762 as a point X-ray source and inferred an AGN based on its power-law index and X-ray luminosity.
Based on the derived [NII]/\Ha\, EL flux ratio  from KMOS (table in M19) and our new [OIII]/\Hb\, EL flux measurement from KMOS YJ-band observations for the central part
of the galaxy 30762, we identify
this object as a type 2 AGN  from the BPT diagram  (right panel of Fig.\,\ref{fig:zHist}).

\subsection{Blob\,2: Extended ionized gas with a velocity gradient, but without a stellar component}
\label{sec:blob2results}

~~~Blob 2 at $z \sim 1.466$  has the widest surface (1270\,kpc$^{2}$) of the three detected blobs (see Table \ref{tab:blobs}) and is the most puzzling blob.
It has no stellar counterpart, shows a velocity rotation pattern  with a velocity gradient of $\sim$300\,km/s,  and lies between two passive galaxies at similar redshift (see Fig.\,\ref{fig:Blob2}).

~~~Using the relation of \citet{ken98}, we inferred an SFR of $1.16\pm 0.30$~M$_\odot$~yr$^{-1}$ from the [OII] EL flux of blob\,2,
assuming that it is related to ionization by young stars.
Blob\,1 (discussed in the previous section) was proved to contain an AGN, so that we cannot use this method to derive its gas mass because its EL flux is not related to ionization by young stars alone.
Using the SFR and the surface of the blob in the Kennicutt-Schmidt relation, we determined the gas surface density,
which allowed us to derive a total gas mass for blob\,2 of  $(1.9 \pm 0.5) \times 10^{10}$~M$_\odot$.  The uncertainties were derived from the uncertainties on the integrated fluxes and the uncertainty on the slope of the equation from \citet{ken98} to transform the [OII] EL flux into the SFR.
They are dominated by the uncertainty on the slope of  the equation from  \citet{ken98}.

The ratio \OIIb/\OIIa\, that is resolved by MUSE can be used to derive the gas density of blob\,2. This line ratio  is $1.4 \pm 0.1$, implying  a gas density $n_e$ $\lesssim$ 10 cm$^{-3}$ \citep{osterbr06},
for the part of the blob that contains the spaxels with the highest [OII] flux (see Fig.\,\ref{fig:Blob2}) and an approaching velocity (blue).
For the rest of the blob with a receding velocity (red in the velocity plot), the ratio \OIIb/\OIIa\, is about one and lower, implying  a higher gas density $n_e$ $\sim$ 100-1000 cm$^{-3}$.

\subsection{Blob\,3: Ionized gas  close to a passive galaxy} 

~~~Blob 3 at $z \sim 1.459$  lies close to a passive galaxy  at similar redshift (see the left panel in Fig.\,\ref{fig:Blob3}). It has only a very faint small stellar counterpart, with a diameter of only 0.3\,arcsec
that corresponds to a surface of about 20\,kpc$^{2}$.
The blob has a surface of 571\,kpc$^{2}$, which is about 30 times larger than the surface of the faint stellar counterpart. It shows a velocity gradient from north to south of about 120\,km/s, with an extended filamentary tail to the south with an extent of  $\sim$10\,kpc.

~~~Using the procedure described in Sect.\,\ref{sec:blob2results} , we inferred an SFR for the entire blob of $1.08\pm 0.28$~M$_\odot$~yr$^{-1}$ from the [OII] EL flux, assuming that it is related to ionization by young stars. The resulting total gas mass is  $(1.4 \pm 0.4) \times 10^{10}$~M$_\odot$.

\subsection{Other galaxies}
\label{sec:31863results}

~~~We detected [OII] in the integrated spectra of all 18 SF cluster galaxies with known spectroscopic redshifts in the MUSE area (see Fig.\,\ref{fig:clustcen}) and for the three AGN.  Given the  MUSE integration time of less than two hours, the S/N per spaxel of the [OII] emission is often quite low,  preventing kinematics studies of about half of the SF XMM2215 cluster galaxies. We present the kinematic diagrams in the appendix \ref{a1} for 9 SF galaxies and two type 2 AGN (object 32152 identified as a type 2 AGN by M19, and object 31746 identified as a type 2 AGN in the BPT diagram Fig.\,\ref{fig:zHist}).
These objects show a regular rotation pattern and a quite regular ionized gas distribution that coincides with the galaxy stellar component, with the exception of objects 38768 and 39683 (Figs.\,\ref{fig:38768} and \ref{fig:39683}), but for less reliable detections at lower S/N for these two objects.

~~~On the other hand, galaxy 31863 at the projected distance of 185\,kpc northwest of the cluster X-ray center
has a more particular kinematic map (see Fig.\,\ref{fig:31863}).  
Object 31863 is an SF galaxy at $z \sim 1.452$ with measurements of several ELs with KMOS and MUSE. In the BPT diagram, it lies in the AGN/SF composite region (see Fig.\,\ref{fig:zHist}).
Its molecular gas mass derived from ALMA observations is $3.5 \pm 0.5 \times 10^{10}$\msun\, \citep{hayashi18}. Fig.\,\ref{fig:31863} shows an elongated [OII] emission region that extends on either side of the galaxy in a filamentary structure with a projected length of about 60\,kpc on each side. In addition to the central stellar counterpart representing galaxy 31863, there is a smaller stellar counterpart in the northern part of the elongated [OII] gas emission, but no stellar counterpart is seen in the HST image for the southern part of the ionized gas.

\begin{table*}[!t]
\caption{Properties of the three detected [OII] blobs. The integrated [OII] flux is given  in units of $10^{-17}$ erg s$^{-1}$~cm$^{-2}$.}
\label{tab:blobs}
\begin{tabular}{ccccccc}
\hline\hline      
Id & RA & DEC & Redshift & Surface & [OII] & Peculiarities\\
\hline
Blob1 & 22:15:58.843 & -17:38:09.98 & 1.453 & 1145 kpc$^{2}$  & $23.3 \pm 3.6$ &X-ray AGN\\
Blob2 & 22:15:58.440 & -17:38:08.64 & 1.466 & 1270 kpc$^{2}$  & $7.1 \pm 1.8$ &no stellar component \\
Blob3 & 22:15:58.972 & -17:38:02.04 & 1.459 & 571 kpc$^{2}$   & $5.9 \pm 1.5$  &close to a passive galaxy\\
\hline                                                                         
\end{tabular}
\end{table*}

The ratio \OIIb/\OIIa\, can be used to derive the gas density of the galaxy 31863. The typical line ratio measured
for the gas around the central stellar counterpart (galaxy 31863) is about 1.5, implying  a gas density $n_e$ $\sim$ 10 cm$^{-3}$ \citep{osterbr06}. A similar \OIIb/\OIIa\, ratio is measured for the southern part of the gas, which has a receding velocity of the gas of about 200\,km/s (red in the velocity map) and lacks a stellar counterpart. For the northern part of the gas
with an approaching velocity  of about 200\,km/s (blue color), the ratio \OIIb/\OIIa\, is lower,  $0.8 \pm 0.2$. This implies  a higher gas density $n_e$ $\sim$ 400-1000 cm$^{-3}$ for the northern part of the gas, which also has a stellar counterpart, but this is smaller than galaxy 31863. 
The typical recombination time of the ionized gas can be calculated as
$\tau_\mathrm{rec} = 1/n_e \alpha_A$,
where $\alpha_A$ is the total recombination coefficient $\alpha_A$ = 4.2 $\times$ 10$^{-13}$ cm$^3$ s$^{-1}$ \citep{osterbr06}. For $n_e$ = 10 cm$^{-3}$, this results in $\tau_\mathrm{rec}$ $\simeq$ 10$^4$\,yr, and for $n_e$ = 400-1000  cm$^{-3}$, this implies $\tau_\mathrm{rec}$ $\simeq$ $100-250$\,yr.

\section{Interpretation}
\label{sec:inter}

\subsection{Blob\,1: Active AGN 30762 becoming the BCG of the cluster?}
\label{sec:obj30762}

~~~The ionized gas of  blob\,1 shows two filamentary patterns, one pattern in the north,  and the other in the northwest. These filamentary patterns of the ionized gas are indicative of gas  outflow with a projected velocity of $\sim 100$\,km/s and $\sim 175$\,km/s, respectively. Since there is no ALMA molecular gas detection for this galaxy and because the gas metallicity \citep[cf.][]{berg21} derived by M19 (albeit in the central region) for object 30762 is high, about solar, inflow of gas in the two filamentary regions is probably unlikely, and the two filamentary structures probably trace the gas outflow.

~~~On its journey through the cluster, galaxy 30762, which hosts an AGN, may have interacted with other SF galaxies  and may have expelled gas from them. This might have produced the [OII] blobs. We describe a possible scenario, whose correctness can be verified by new future simulations of forming clusters around cosmic noon,  following galaxies that travel through the cluster, including warm gas such as the [OII] ionized gas we detected.

~~~In the past, moving through the cluster, AGN 30762 may have interacted with galaxy 781 near blob\,2 and may have expelled gas from galaxy 781, which is passive today \citep{beifiori17}. Due to proper motions of galaxies in the cluster, this might have occurred at an apparent redshift  of $z \sim 1.460$. We note that blob\,1 (blue ellipse) and blob\,2 (larger red ellipse) are close to each other and even slightly overlap partly in projection (see Fig.\,\ref{fig:clustcen}). Continuing its journey into our line-of-sight direction, AGN 30762 now appears due to its proper motion at redshift $z \sim 1.453$, while the outflowing receding gas flew in the oposite direction through the cluster and reached redshift $z \sim 1.466$. We see the projection of the gas outflowing as blob\,2 with a proper motion of the gas, inducing a projected line-of-sight velocity difference of about 300~km~s$^{-1}$
([OII] velocity map in Fig.\,\ref{fig:Blob2}).
The gas in blob\,2 might either be due to a reservoir fed by outflows from the AGN 30762 or to gas stripped from galaxy 781 during the interaction with AGN 30762 and gas stripped  from galaxy 781 due to  ram-pressure stripping (RPS) in the inner region of the cluster, which was then trapped by the gravitational potential of blob\,2.
This would agree with the AGN-RPS synergy quenching scenario advocated by \citet{maier22} for the inner regions of clusters at intermediate redshifts.

~~~Galaxy 30762 hosts an AGN and lies at a projected distance of 74\,kpc from the cluster X-ray center (tiny open black circle in Fig.\,\ref{fig:clustcen}). It is a good candidate for becoming the (active) brightest cluster galaxy (BCG) of the cluster XMM2215 if it reaches the cluster X-ray center in the future \citep[cf., e.g., simulations by][]{muldrew18}. With a velocity of $\sim$1000\,km/s (as derived by  M19), which corresponds to about  1\,kpc/Myr, it could reach the cluster X-ray center in about 0.1\,Gyr.

\subsection{Blob\,2: Outflowing gas, stripped gas, or shocks?}
\label{sec:blob2int}

~~~Blob 2 at $z \sim 1.466$  has no stellar counterpart, shows a velocity rotation pattern,  and lies between two passive galaxies at similar redshift (see Fig.\,\ref{fig:Blob2}).
The typical line ratio measured pixel per pixel  for the east part  with an approaching velocity (blue in the velocity map) is about 1.5, implying  a gas density $n_e$ $\lesssim$ 10 cm$^{-3}$ \citep{osterbr06}. This part of the blob has the highest [OII] emission and highest velocity dispersion, as shown in Fig.\,\ref{fig:Blob2}. The total recombination coefficient for this density is $\tau_\mathrm{rec}$ $\simeq$ 10$^4$ yr. This quite short recombination time  implies that  a source of gas excitation must be present in the gas component. Shocks might be an explanation, as also suggested by the high velocity dispersion in this region.

~~~The high mass of ionized gas of blob\,2 of $(1.9 \pm 0.5) \times 10^{10}$~M$_\odot$ derived in Sect.\,\ref{sec:blob2results} and the absence of stars indicate that
gas might be stripped from galaxy 781 during interactions or merger events, and it might be trapped by the gravitational potential of the structure, like in the scenario discussed in Sect.\,\ref{sec:obj30762}.
This gas mass  is typical for SF galaxies with stellar masses of a few $10^{10}$~M$_\odot$ \citep[e.g.,][]{cyb17}, like the passive galaxy 781 in the vicinity of blob\,2 and at a similar redshift.
Blob\,2 lies at a projected distance of about 70\,kpc from the cluster X-ray center. The central part of the cluster with a diameter of 200\,kpc (inner black circle in Fig.\,\ref{fig:clustcen}) contains six passive galaxies with known spectroscopic redshifts \citep{hilton10,beifiori17,hayashi18} and the three blobs. This inner area of the cluster with a radius of 100\,kpc is a
region in which SF galaxies stop their star formation, probably due to different processes. Likely candidates are mergers, RPS, and shocks.

~~~We are currently not aware of any simulations of cool gas (T $\sim 10 000$K) in massive galaxy clusters near cosmic noon. Very recently, a first step in this direction was made by simulations of cool gas in galaxy clusters  at redshift zero \citep{staff25}. The authors combined the TNG-Cluster and TNG300 cosmological magnetohydrodynamical simulations to study cool gas in local galaxy clusters. If these theoretical studies of cool gas in clusters were extended to massive clusters at higher redshifts (near cosmic noon), our slightly speculative scenarios for the [OII] blobs discovered in XMM2215 might be verified.

\subsection{Blob\,3: Ionized gas stripped by RPS or shocks?} 

~~~Blob\,3 at $z=1.459$ is located near the passive galaxy 39468 at a projected distance of 62\,kpc from the  cluster X-ray center. This ionized gas  might be caused by RPS acting on the galaxy 39468 in the past when the galaxy entered the inner region of the cluster.
As mentioned before, the central part of the cluster with a diameter of 200\,kpc contains six passive galaxies (including 39468),
and in this 
region of the cluster star formation in galaxies is quenched.
The very  faint stellar counterpart in the center of blob\,3 might be star formation induced in the stripped gas, as observed during RPS of gas at lower redshifts.

~~~ Assuming that it is related to ionization by young stars, we inferred a total gas mass of  $(1.4 \pm 0.4) \times 10^{10}$~M$_\odot$ for this blob. However, given the faintness of the stellar counterpart, it is quite unlikely that this is the only source for ionization of blob\,3. Therefore, additional ionization due to shocks from past and ongoing merger events might be an explanation.

~~~While the photoionization of blob\,1 is related to the AGN in galaxy 30762, the photoionization of blob\,2 and blob\,3  might have diverse origins, and it remains enigmatic. Galaxy interactions may induce shocks through the gas flows that are caused by the merger process \citep{rich15}. Photoionization can be powered by an AGN as well, which can be fuelled with gas that is driven inward by tidal forces during mergers.
The galaxy velocity distribution in cluster XMM2215 is bimodal (Figs.\,\ref{fig:clustcen} and \ref{fig:zHist}), so that mergers and shocks are very probable.
Nevertheless, the enigmatic origin of the photoionization of blob\,2 and blob\,3
has also to be addressed with future improved simulations of massive clusters that form around cosmic noon, including not only the hot gas component, but also warm gas, such as the [OII] ionized gas we detected.

\subsection{Object 31863: Merger event}

~~~Galaxy 31863 is an SF galaxy at $z \sim 1.452$ at a projected distance of 185\,kpc
from the cluster X-ray center.
It has an elongated [OII] emission region that extends on either side of the galaxy (see Fig.\,\ref{fig:31863}) in a filamentary structure with a projected length of about 60\,kpc on each side. The density analysis in Sect.\,\ref{sec:31863results} implies a gas density of $n_e$ $\sim$ 400-1000 cm$^{-3}$ for the northern part of the gas, which is higher than in the central and southern part. The typical recombination time of the ionized gas  for this density  is $\tau_\mathrm{rec}$ $\simeq$ $100-250$\,yr.

~~~The very short recombination time of $\simeq$ $100-250$\,yr for the northern part of the gas implies that  a source of gas excitation must be present in the
gas. A smaller stellar counterpart for the northern part is visible in the HST image, and we call this smaller galaxy 31863b.
The southern part of the gas has a longer recombination time of $\tau_\mathrm{rec}$ $\simeq$ $10^{4}$\,yr and no stellar counterpart.  It most likely represents an outflow of gas that is produced by a central starburst fed by the collapse of gas induced by the merging process of galaxies 31863 and 31863b. The brightest [OII] regions seen in the central part of the gas support this scenario of a central starburst.

\subsection{Caveats and next steps}
\label{sec:caveats}

~~~The BPT line diagnostic diagram is useful for constraining sources for photoionization, for instance, the presence of shocks. However, only five galaxies from our studied sample have the all EL measurements that are required for the BPT diagram (see Fig.\,\ref{fig:zHist}, right panel).  
The reason that only very few objects  (5 out of the 19 galaxies studied by M19) have measurements that can be plotted in the BPT diagram is that 
at least one of the four ELs ([NII]$\lambda$6584, H$\alpha$, [OIII]$\lambda$5007, or H$\beta$) that are needed to explore the ionizing conditions in the BPT diagram at $z\sim 1.5$ is heavily affected by strong night-sky lines (see also table in M19) when near-infrared  spectroscopy at $1-2\mu\rm{m}$ from ground-based telescopes such as VLT is employed. This shows the absolute need for observations with the James Webb Space Telescope (JWST), which alone of the current telescopes is able to provide measurements of all four EL fluxes free from strong night-sky lines for a statistically significant number of cluster galaxies at these redshifts.

~~~The integration time for the MUSE data was shorter than two hours. Although enigmatic and interesting blobs with brighter [OII] ELs could be identified and also used for kinematics, no kinematic studies were possible because the ELs are too faint and the S/N of the [OII] flux in single spaxels is too low for about half of the  SF cluster galaxies observed with MUSE. 
Therefore, deeper MUSE observations would be an asset for studying the kinematics of all known spectroscopic SF cluster members in the inner part of XMM2215.

\section{Summary and conclusions}
\label{sec:disc}
~~~Using VLT-MUSE  integral field spectroscopy, we investigated the central 500$\times$500\,kpc$^{2}$  area of the XMM2215 cluster at $z \sim 1.46$. This area contains 28 spectroscopically identified cluster galaxies. Our analysis was complemented with VLT-KMOS observations for some objects and with other archival ancillary data. Our principal findings are summarized below.

\begin{enumerate}
\item
  We detected \OIId\, emission lines in the integrated spectra of 21 galaxies. The remaining 7 galaxies are passive galaxies.  The central part of the cluster has a diameter of 200\,kpc (inner black circle in Fig.\,\ref{fig:clustcen}) and contains 6 passive galaxies. In this
region, star formation in galaxies is quenched.
\item
The galaxy velocity distribution in this high-redshift cluster is bimodal  (see Figs.\,\ref{fig:clustcen} and \ref{fig:zHist}). This indicates that the cluster is probably not fully virialized, and ongoing merging events might develop shocks that might be a source of photoionization. 
\item
We discovered three diffuse ionized \OIId\, gas structures that we called [OII] blobs. They lie
in the central
area of the cluster within a radius of 100\,kpc.  
\item
Blob\,1 with its stellar counterpart galaxy 30762 at $z \sim 1.453$ has a surface of 1145\,kpc$^{2}$ and lies at a projected distance of 74\,kpc from the cluster X-ray center (tiny open black circle in Fig.\,\ref{fig:clustcen}).  The source of ionization of the blob is an AGN, and the galaxy is detected in X-rays and is classified  as a type 2 AGN from the BPT diagram. The blob displays two prominent filamentary patterns that indicate outflow of gas  with a projected velocity of $\sim 100$\,km/s and $\sim 175$\,km/s (see Fig.\,\ref{fig:AGN30762}). This galaxy might very likely become the BCG of the cluster in the near future. 
\item
Blob\,2  at $z \sim 1.466$ with a surface of 1270\,kpc$^{2}$ is the most puzzling of the three blobs.  It has no stellar counterpart, shows a velocity rotation pattern  with a velocity gradient of $\sim$300\,km/s,  and lies between two passive galaxies at similar redshift  (see Fig.\,\ref{fig:Blob2}). We discussed a possible formation and photoionization scenario, according to which AGN 30762 traveled through the cluster and interacted with galaxy 781, which lies close to blob\,2. Additionally, shocks from past and ongoing merger events indicated by the bimodal velocity distribution in this cluster might be a source of photoionization.
\item
  Blob\,3  at $z \sim 1.459$ has a surface of 571\,kpc$^{2}$ and is located near the passive galaxy 39468 (see Fig.\,\ref{fig:Blob3})
  at a projected distance of 62\,kpc from the cluster X-ray cluster. Our interpretation is that the blob is the result of a mix of RPS acting on galaxy 39468 in the past when the galaxy entered the inner region of the cluster and shocks  from past and ongoing merger events, as indicated by the bimodal velocity distribution in this cluster. 
\item
  We also analyzed galaxy 31863, which lies at a projected distance of 185\,kpc northwest of the cluster X-ray center and has a more particular kinematic map. The density and  recombination-time analysis for this object  revealed that the stellar counterpart in the northern part of the gas must be the source of gas excitation. It represents the smaller galaxy 31863b, which is in a merging process with galaxy 31863. We interpret the southern part of the gas as outflow of gas that is produced by the central starburst, which is fed by the collapse of gas induced by the merging process.
\end{enumerate}

~~~To extend this analysis to more cluster galaxies and to further explore the photoionization sources of the enigmatic blobs\,2 and 3, JWST observations are required to overcome night-sky-line contamination of the EL fluxes that are required for the BPT diagram, as well as deeper MUSE observations. We  used MUSE observations
with an integration time shorter than two hours for this work. To further explore the origin of the enigmatic blobs, better simulations of forming clusters around cosmic noon are also needed that include not only the hot-gas component, but also warm gas. This needs to be compared with the warm ionized gas blobs we observe in XMM2215.

\begin{acknowledgements}
We would like to thank the anonymous referee for providing constructive comments and help in improving the manuscript.
TK acknowledges financial support from JSPS KAKENHI Grant Number 24H00002 (Specially Promoted Research by T. Kodama et al.)
\end{acknowledgements}

\clearpage

\begin{appendix}
\section{EL flux and velocity maps}
\label{a1}

\begin{figure}[!h]
  \centering
  \captionsetup{width=0.45\textwidth}
    \includegraphics[width=0.45\textwidth,angle=0,clip=true]{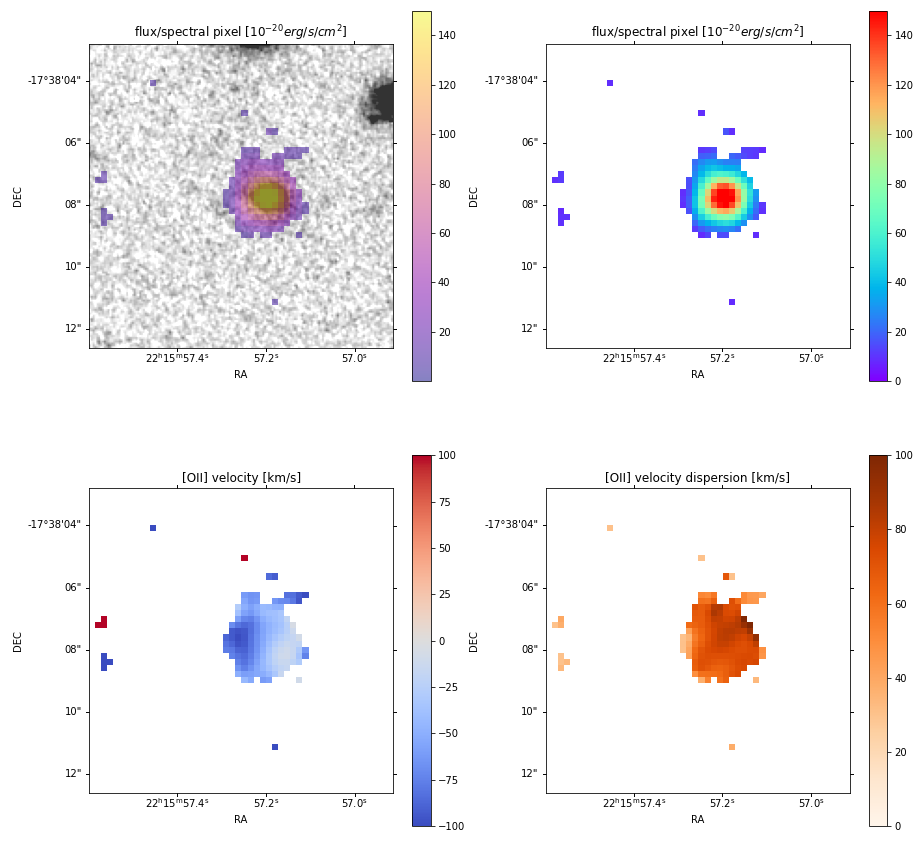}
   \centering   
   \caption{\footnotesize Clockwise starting in the upper left panel:  HST F160W image around the galaxy 31179 with overplotted [OII] EL map; spatially resolved [OII] EL map; spatially resolved [OII] velocity dispersion map;  spatially resolved [OII] velocity map. The white background corresponds to spaxels with $\rm{S/N_{[OII]}<5}$.
   }  
\label{fig:31179}
\end{figure}

\begin{figure}[!h]
  \centering
  \captionsetup{width=0.45\textwidth}
    \includegraphics[width=0.45\textwidth,angle=0,clip=true]{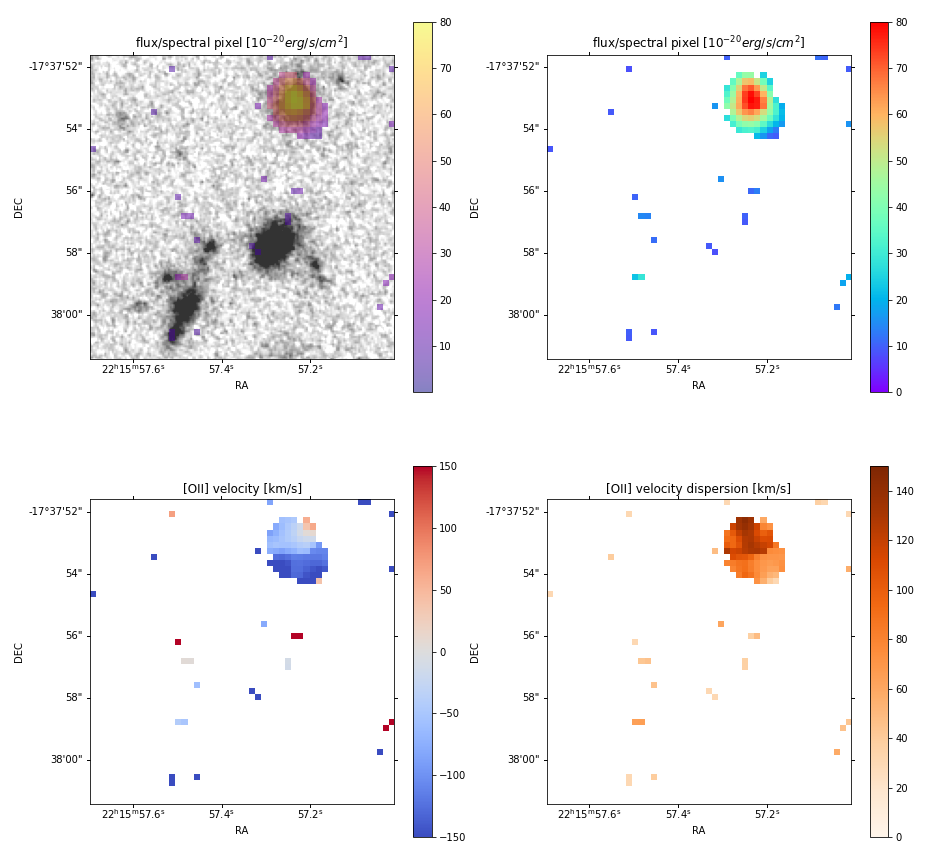}
   \centering   
   \caption{\footnotesize Clockwise starting in the upper left panel:  HST F160W image including the galaxy 31680 in the north-west corner of the image and three other galaxies with fainter [OII], with overplotted [OII] EL map; spatially resolved [OII] EL map; spatially resolved [OII] velocity dispersion map;  spatially resolved [OII] velocity map. The white background corresponds to spaxels with $\rm{S/N_{[OII]}<5}$. From ALMA, galaxy 31680 has a derived molecular gas mass of $8.1^{+0.6}_{-0.5} \times 10^{10}$\msun\, \citep{hayashi18}. Two other galaxies in the figure have lower ALMA molecular gas mass measurements of $(2.7\pm 0.6) \times 10^{10}$\msun\,and $(5.8\pm 0.6) \times 10^{10}$\msun, but they are not detected in [OII] in the individual spaxels above the S/N cut limit.
   } 
\label{fig:31680}
\end{figure}

\begin{figure}[!h]
  \centering
  \captionsetup{width=0.45\textwidth}
    \includegraphics[width=0.45\textwidth,angle=0,clip=true]{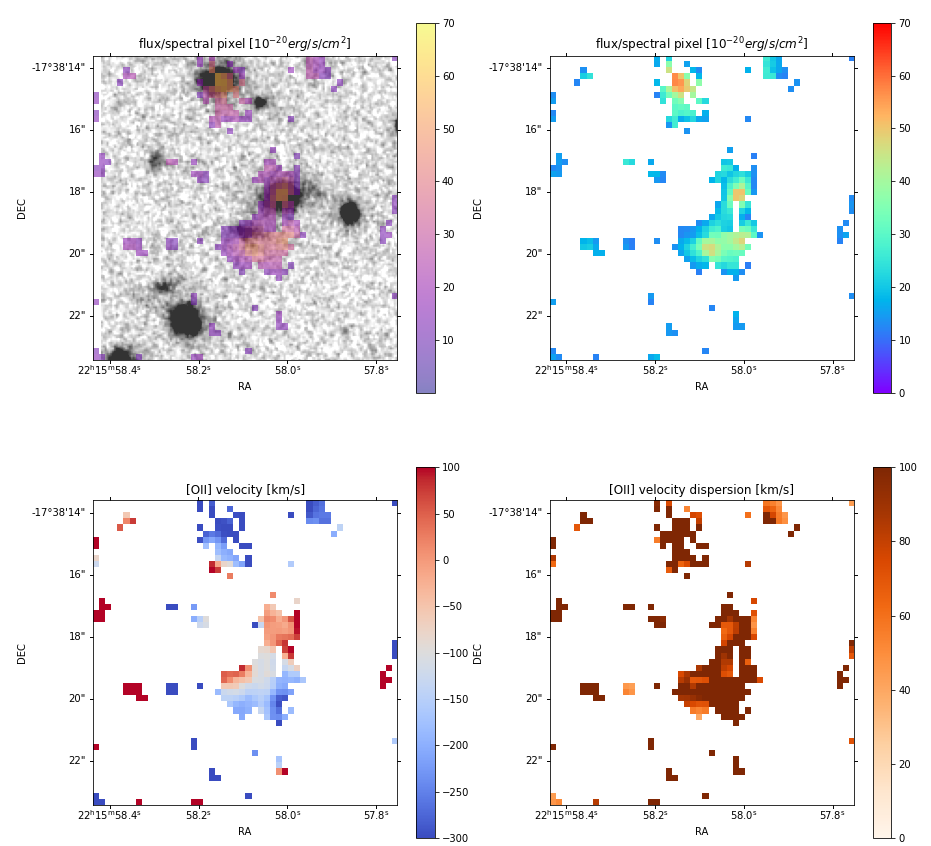}
   \centering   
   \caption{\footnotesize Clockwise starting in the upper left panel:  HST F160W image including the galaxy 38910 in the north-east upper part of the image and 38768 in the center, with overplotted [OII] EL map; spatially resolved [OII] EL map; spatially resolved [OII] velocity dispersion map;  spatially resolved [OII] velocity map. The white background corresponds to spaxels with $\rm{S/N_{[OII]}<3}$. From ALMA, galaxy 38768 has a derived molecular gas mass of $(3.1 \pm 0.6) \times 10^{10}$\msun\, \citep{hayashi18}.
   } 
\label{fig:38768}
\end{figure}

\begin{figure}[!h]
  \centering
  \captionsetup{width=0.45\textwidth}
    \includegraphics[width=0.45\textwidth,angle=0,clip=true]{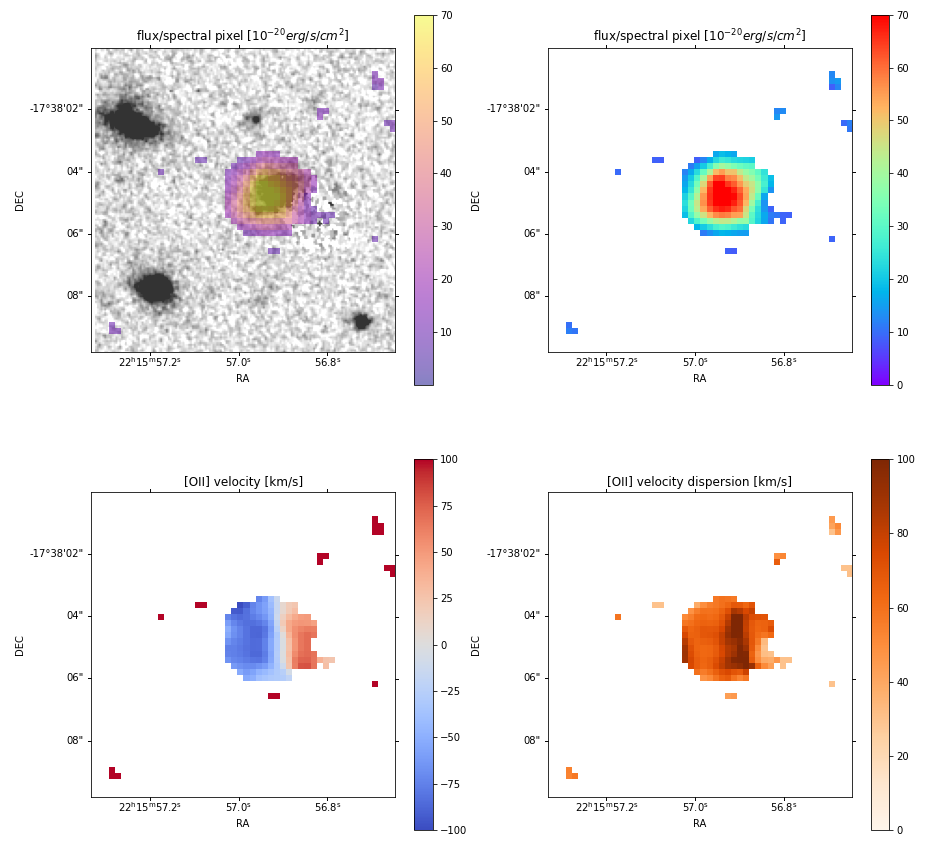}
   \centering   
   \caption{\footnotesize Clockwise starting in the upper left panel:  HST F160W image around the galaxy 31150 with overplotted [OII] EL map; spatially resolved [OII] EL map; spatially resolved [OII] velocity dispersion map;  spatially resolved [OII] velocity map. The white background corresponds to spaxels with $\rm{S/N_{[OII]}<5}$.
   } 
\label{fig:31150}
\end{figure}

\begin{figure}[!h]
  \centering
  \captionsetup{width=0.45\textwidth}
    \includegraphics[width=0.45\textwidth,angle=0,clip=true]{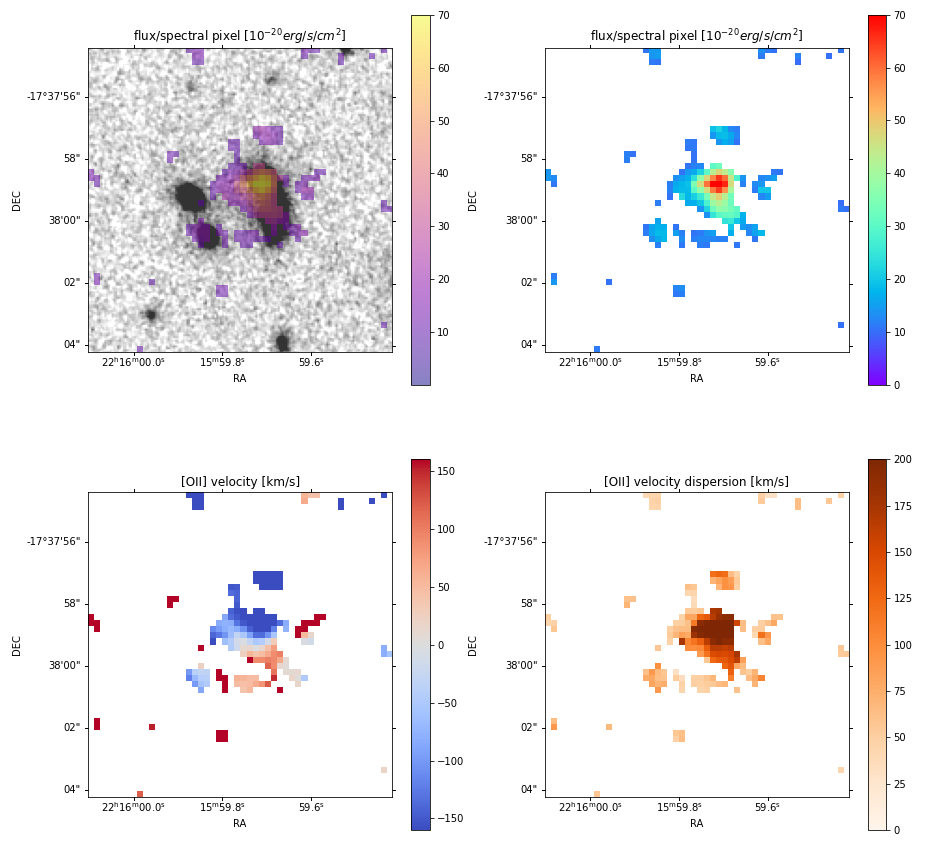}
   \centering   
   \caption{\footnotesize Clockwise starting in the upper left panel:  HST F160W image around the galaxy 39683 with overplotted [OII] EL map; spatially resolved [OII] EL map; spatially resolved [OII] velocity dispersion map;  spatially resolved [OII] velocity map. The white background corresponds to spaxels with $\rm{S/N_{[OII]}<3}$. From ALMA, galaxy 39683 has a derived molecular gas mass of $10.5^{+0.7}_{-0.5} \times 10^{10}$\msun\, \citep{hayashi18}.
   } 
\label{fig:39683}
\end{figure}

\begin{figure}[!h]
  \centering
  \captionsetup{width=0.45\textwidth}
    \includegraphics[width=0.45\textwidth,angle=0,clip=true]{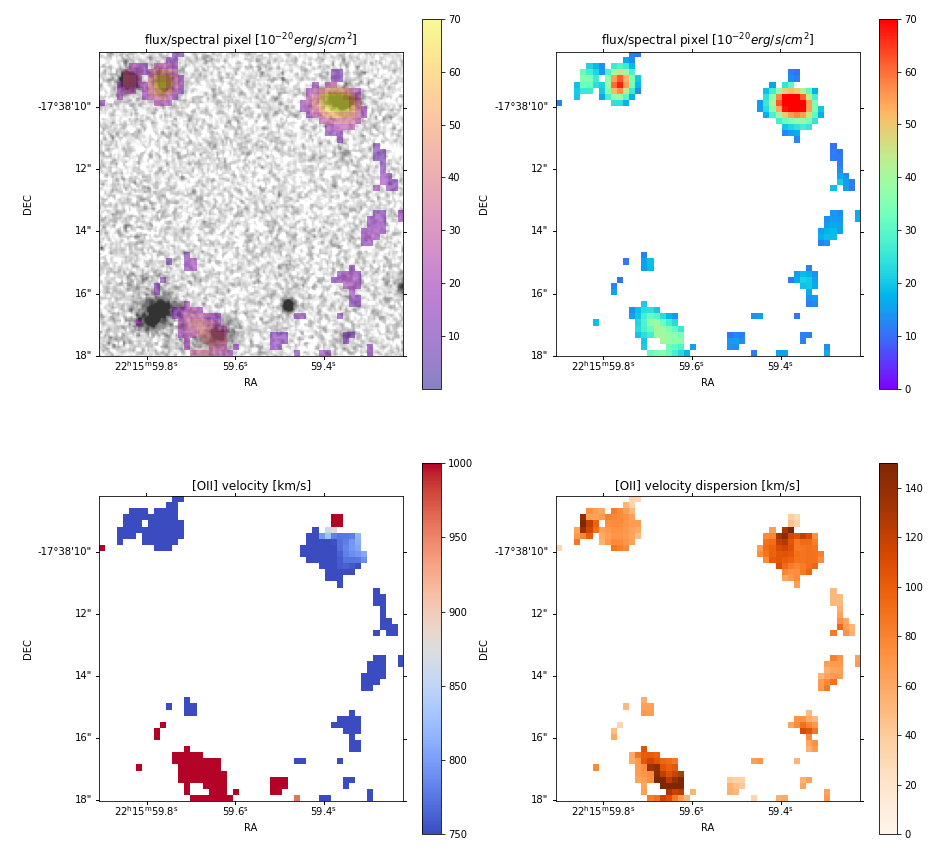}
   \centering   
   \caption{\footnotesize Clockwise starting in the upper left panel:  HST F160W image including the galaxy 823 in the north-west upper part of the image, object 13 in the south part and another galaxy at slightly lower redshift in  the north-east upper part of the image, with overplotted [OII] EL map; spatially resolved [OII] EL map; spatially resolved [OII] velocity dispersion map;  spatially resolved [OII] velocity map. The white background corresponds to spaxels with $\rm{S/N_{[OII]}<4}$.
   } 
\label{fig:31150}
\end{figure}

\begin{figure}[!h]
  \centering
  \captionsetup{width=0.45\textwidth}
    \includegraphics[width=0.45\textwidth,angle=0,clip=true]{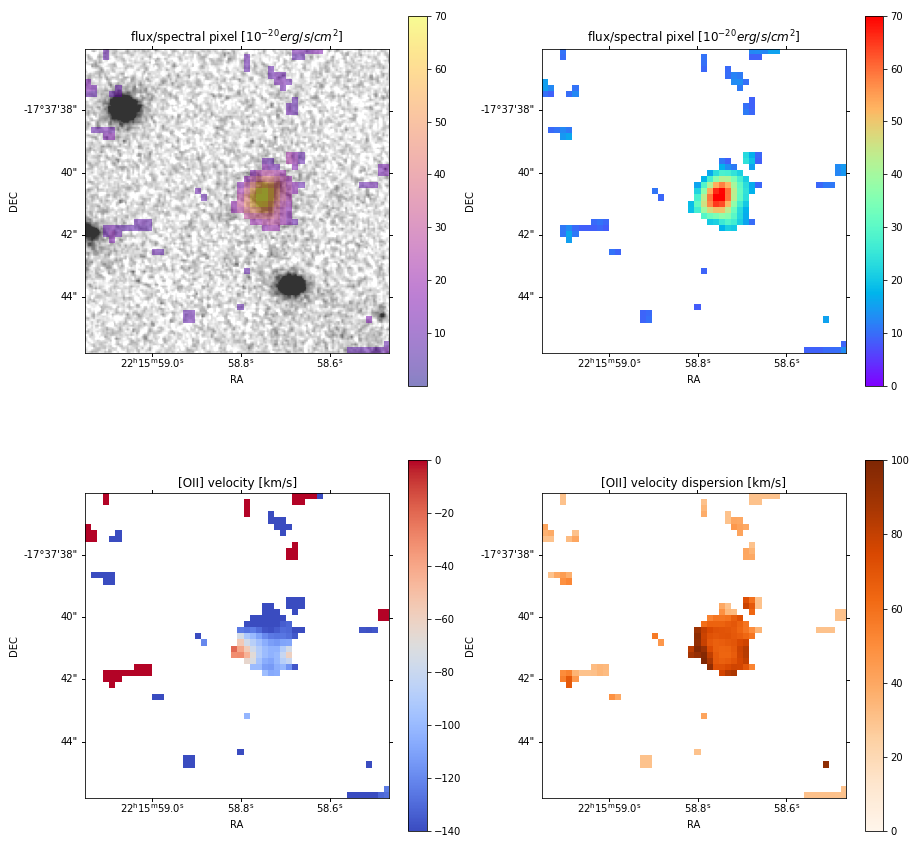}
   \centering   
   \caption{\footnotesize Clockwise starting in the upper left panel:  HST F160W image around the galaxy 32152 hosting a type-2 AGN with overplotted [OII] EL map; spatially resolved [OII] EL map; spatially resolved [OII] velocity dispersion map;  spatially resolved [OII] velocity map. The white background corresponds to spaxels with $\rm{S/N_{[OII]}<4}$.  From ALMA, galaxy 32152 has a derived molecular gas mass of $6.6^{+0.8}_{-0.9} \times 10^{10}$\msun\, \citep{hayashi18}.
   } 
\label{fig:32152}
\end{figure}

\begin{figure}[!h]
  \centering
  \captionsetup{width=0.45\textwidth}
    \includegraphics[width=0.45\textwidth,angle=0,clip=true]{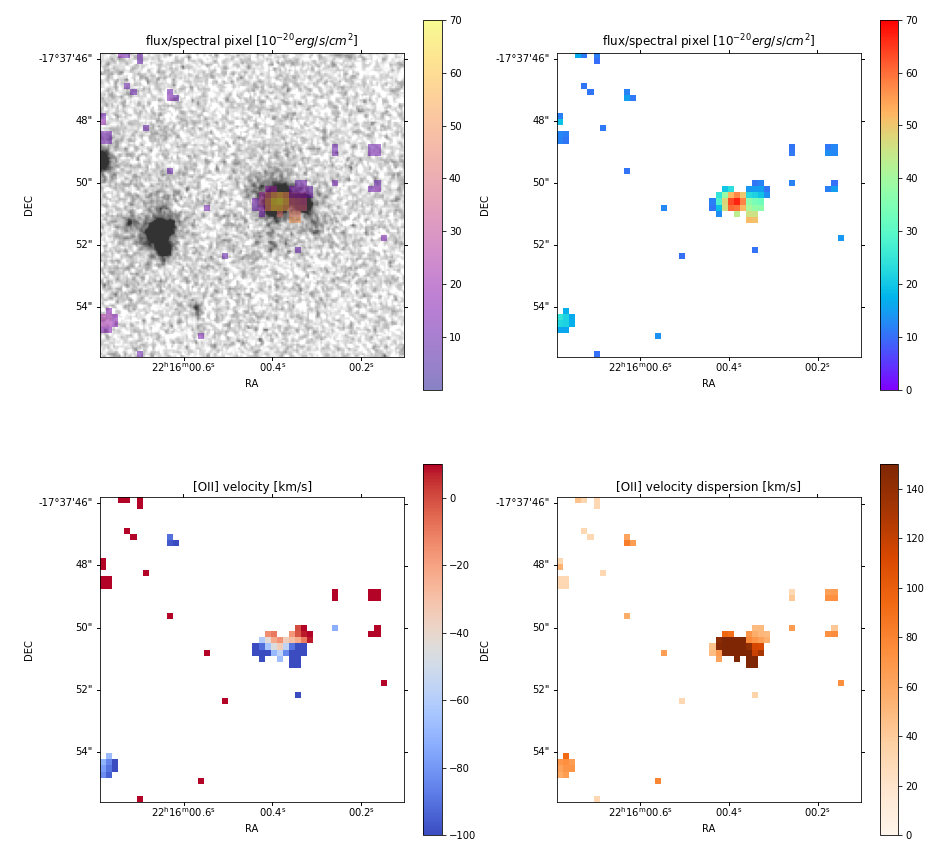}
    \centering
   \caption{\footnotesize Clockwise starting in the upper left panel:  HST F160W image around the galaxy 31746 hosting a type-2 AGN with overplotted [OII] EL map; spatially resolved [OII] EL map; spatially resolved [OII] velocity dispersion map;  spatially resolved [OII] velocity map. The white background corresponds to spaxels with $\rm{S/N_{[OII]}<3}$.  From ALMA, galaxy 31746 has a derived molecular gas mass of $(3.2 \pm 0.6) \times 10^{10}$\msun\, \citep{hayashi18}.
   } 
\label{fig:31746}
\end{figure}

\end{appendix}

\end{document}